\documentclass[a4paper,11pt]{article}
\usepackage{pos}
\usepackage{hyperref}
\usepackage{setspace}

\title{Design and upgrade of the prototype Schwarzschild-Couder Telescope}
\ShortTitle{Design and Upgrade of the pSCT}

\manuallySeparateAuthors
\author*[a]{L.~P.~Taylor}
\author{on behalf of the CTA SCT Consortium}

\affiliation[a]{Department of Physics and Wisconsin IceCube Particle Astrophysics Center,\\ University of Wisconsin, Madison, WI 53706, USA}

\emailAdd{ltaylor23@wisc.edu}

\abstract{The Cherenkov Telescope Array (CTA) is the next-generation ground-based observatory for very-high energy gamma-ray astronomy. CTA will have unparalleled sensitivity and angular resolution and will detect gamma-ray sources nearly 100 times faster than current arrays, enabling valuable multiwavelength and multimessenger observations. The Schwarzschild-Couder Telescope (SCT) is a candidate for the Medium-Sized telescope in CTA. A prototype SCT (pSCT) has been constructed at the Fred Lawrence Whipple Observatory in Arizona USA. Its camera is currently partially instrumented with 1600 pixels (2.7 degree FOV). The small plate scale of the optical system allows densely packed silicon photomultipliers to be used, which combined with high-density trigger and waveform readout electronics enable the high-resolution camera. The camera's electronics are capable of imaging air shower development with waveform readout with nanosecond resolution. The pSCT was inaugurated in January 2019, with commissioning continuing throughout that year. The first campaign of observations with the pSCT was conducted in January and February of 2020. Gamma-ray emission from the Crab Nebula was detected with a significance of 8.6 sigma. An upgrade to the pSCT camera is currently underway. The upgrade will fully populate the focal plane, increasing the field of view to 8 degrees diameter, and lower the front-end electronics noise, enabling a lower trigger threshold and improved reconstruction and background rejection.}

\FullConference{%
  7th Heidelberg International Symposium on High-Energy Gamma-Ray Astronomy (Gamma2022)\\
  4-8 July 2022\\
  Barcelona, Spain\\}

\begin{document}
\maketitle

\section{CTA and the pSCT}


\begin{figure}[b]
    \centering
    \includegraphics[width=0.41\linewidth]{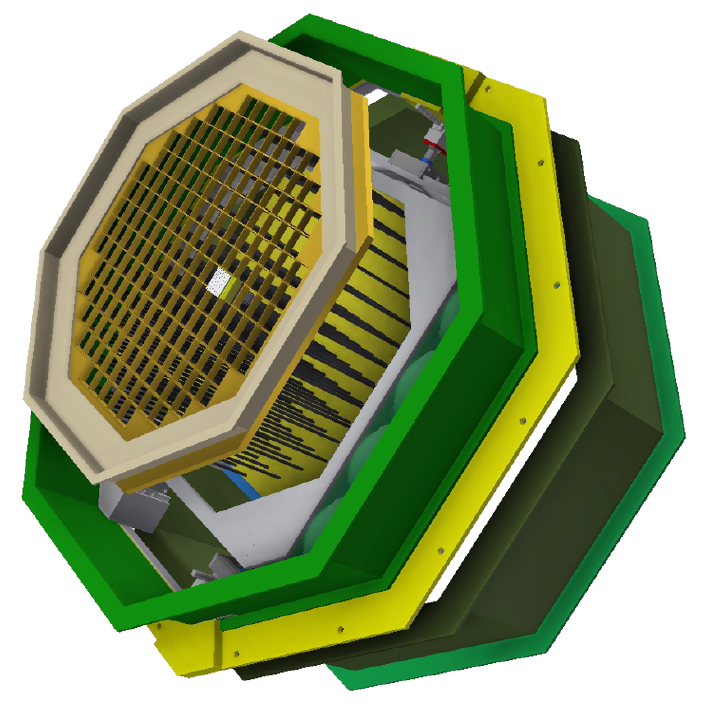}
    \includegraphics[width=0.08\linewidth, angle=90]{Figures/CameraModule.png}
    \caption{\textbf{Top}: exploded front view of the camera mechanical design. Modules are inserted through the front lattice and rest between carbon fiber rods which connect the lattice to the back bulkhead. There are two holes per module in the back bulkhead, one for a backplane connector and one for a securing screw. Enclosing the camera is a shroud (dark green). The camera is secured to the telescope via the outer structure (light green). The inner structure of the camera can be moved relative to the outer structure using three motors. \textbf{Bottom}: exploded side view of a single module. To the left is the FPM including the SIPM, insulating foam and heatsink. The auxiliary (top) and primary (bottom) boards are shown outside of the aluminum FEE housing. The primary board is longer than the auxiliary board because of the backplane connector. \cite{Taylor2019Design}}
    \label{fig:MechanicalDesign}
\end{figure}

When very high energy photons hit the atmosphere they initiate extensive air showers whose constituents move faster than the speed of light in the atmosphere. These constituent particles emit Cherenkov radiation which can be detected at the ground by Imaging Atmospheric Cherenkov Telescopes (IACTs).

The Cherenkov Telescope Array (CTA) is an observatory featuring three telescope sizes (Small, Medium, and Large) each of which will cover different energy bands between 20~GeV and 300~TeV. \cite{acharyya2019monte} The CTA ``alpha configuration'' will have two sites, one in each hemisphere. The northern site is planned to have four Large-Sized telescopes and nine Medium-Sized telescopes, will cover 3~km$^2$, and will be located in La Palma, Spain. The southern site is planned to have 14 Medium-Sized telescopes and 37 Small-Sized telescopes, will cover 0.25~km$^2$, and will be located in Paranal, Chile. Together these sites will cover the whole sky and have and order of magnitude greater sensitivity than other instruments. \cite{consortium2017science}

The prototype Schwarzschild-Couder Telescope (pSCT) is a candidate for the CTA Medium-Sized telescope. While most IACTs use Davies-Cotton (single-mirror parabolic) optics the pSCT will use novel Schwarzchild-Couder (dual-mirror) optics with a 9.66~m primary mirror, a 5.4~m secondary mirror, and a curved focal plane located between them. \cite{CameraPaper} Dual-mirror optics produce an excellent optical point spread function, a wide field of view, and a smaller plate scale when compared to single-mirror optics. \cite{vassiliev2007schwarzschild} Because of the small plate scale (1.625 mm per minute of arc) the pSCT is able to use Silicon photomultipliers (SiPMs) which are smaller than traditional PMTs and thus result an improved image resolution. This improved resolution in turn reduces uncertainty in gamma-ray direction and energy resolution resulting in improved background rejection.

\section{pSCT Camera Design}

\begin{figure}[b]
    \centering
    \includegraphics[width=0.54\linewidth]{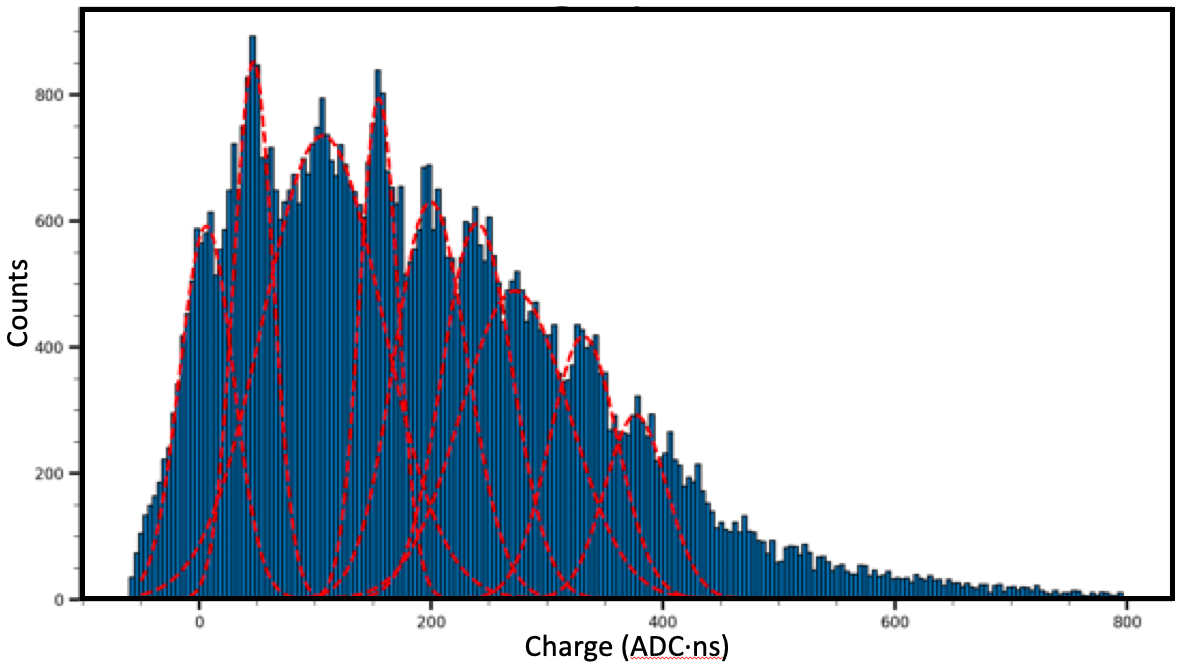}
    \hfill
    \includegraphics[width=0.44\linewidth]{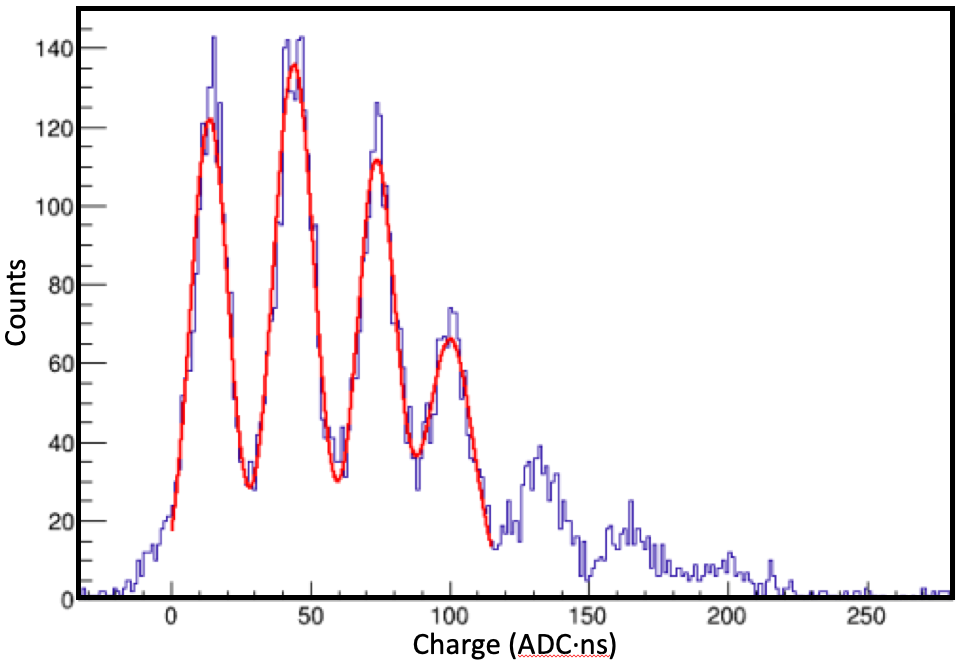}
    \caption{\textbf{Right}: Charge spectrum produced using current camera modules including existing front-end electronics and Hammamatsu SiPMs. \textbf{Left}: Charge spectrum produced using upgraded FBK SiPMs and upgraded front-end electronics. The upgraded modules have lower noise and show a significant improvement in charge spectrum resolution.}
    \label{fig:ChargeSpectrum}
\end{figure}

The pSCT camera is comprised of an outer structure (which is attached to the telescope) and an inner structure (which is designed to hold the camera modules). This mechanical design is shown in figure \ref{fig:MechanicalDesign}. The inner structure contains a front aluminum lattice and back bulkhead which are connected with internal carbon fiber rods. Modules are inserted through the front lattice and secured at the back bulkhead. Backend electronics are secured to the other side of the back bulkhead. The inner structure (including modules) can be moved relative to the outer camera structure via a camera alignment system. This allows the focal plane to be positioned in situ at the correct distance to the secondary mirror. \cite{adams2020verification} The final pSCT design will include 177 modules arranged in a roughly circular shape and nine backplanes, each capable of handling up to 25 modules. Currently the pSCT is instrumented only with the central 25 modules resulting in a 2.7$^{\circ}$ square field of view.

Each module contains Front End Elecronics (FEE) as well as a focal-plane module (FPM) which contains 64 image pixels. Two types of photosensor are used in the current camera: third-generation near-ultraviolet high-density SiPMs (NUV-HD3) which were produced by Fondazione Bruno Kessler (FBK) in collaboration with Istituto Nazionale di Fisica Nucleare (INFN), and Hamamatsu S12642-0404PA-50(X) photon detectors. Pixels are grouped into quadrants which are mounted onto a printed circuit board with a tapped copper block on the other side. The quadrants are bolted into the top of a copper post which passes through a layer of insulating foam and is secured at its base by a plastic base plate. Module FEEs generally handle readout and control of the FPM. FEE main functionalities are amplification and digitization of FPM signals, temperature monitoring, module-level trigger generation, and waveform data packaging and transfer to storage. \cite{CameraPaper} 

Modules are inserted into the camera lattice from the front and secured via two holes in the back bulkhead. One hole is for the module connector which connects each module to the backend electronics. The second hole is for a securing screw which holds the modules and backend electronics in place on the bulkhead. Once installed the focal-plane modules together form a curved focal plane which faces the secondary mirror.

The backend electronics are responsible for power supply management and housekeeping, data routing, backplane-level trigger generation, camera module synchronization, and event time-tagging. The event data stream which is generated by the backend electronics is sent to a data server close to the telescope which provides slow-control and run-control software. slow-control software handles powering and monitoring hardware components of the camera. The main Run-control functions are to load the FEE configuration settings prior to data-acquisition, to start, monitor, and stop runs, and to manage the recording of data to disk.\cite{CameraPaper} 

A shutter and heat management system protect the camera from the elements and provide a stable operating temperature. Flashers are currently installed on an optical table at the center of the secondary mirror in order to provide the camera with consistent triggers at a known rate.

\section{Camera Upgrade}

\begin{figure}[t]
    \centering
    \includegraphics[width=0.375\linewidth]{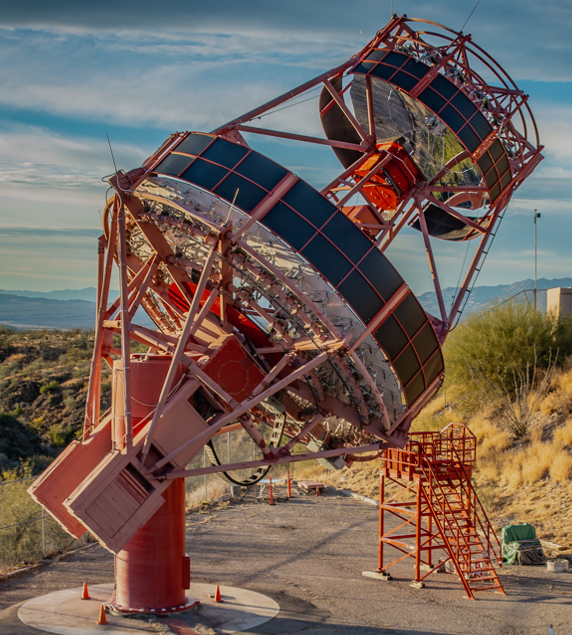}
    \hfill
    \includegraphics[width=0.61\linewidth]{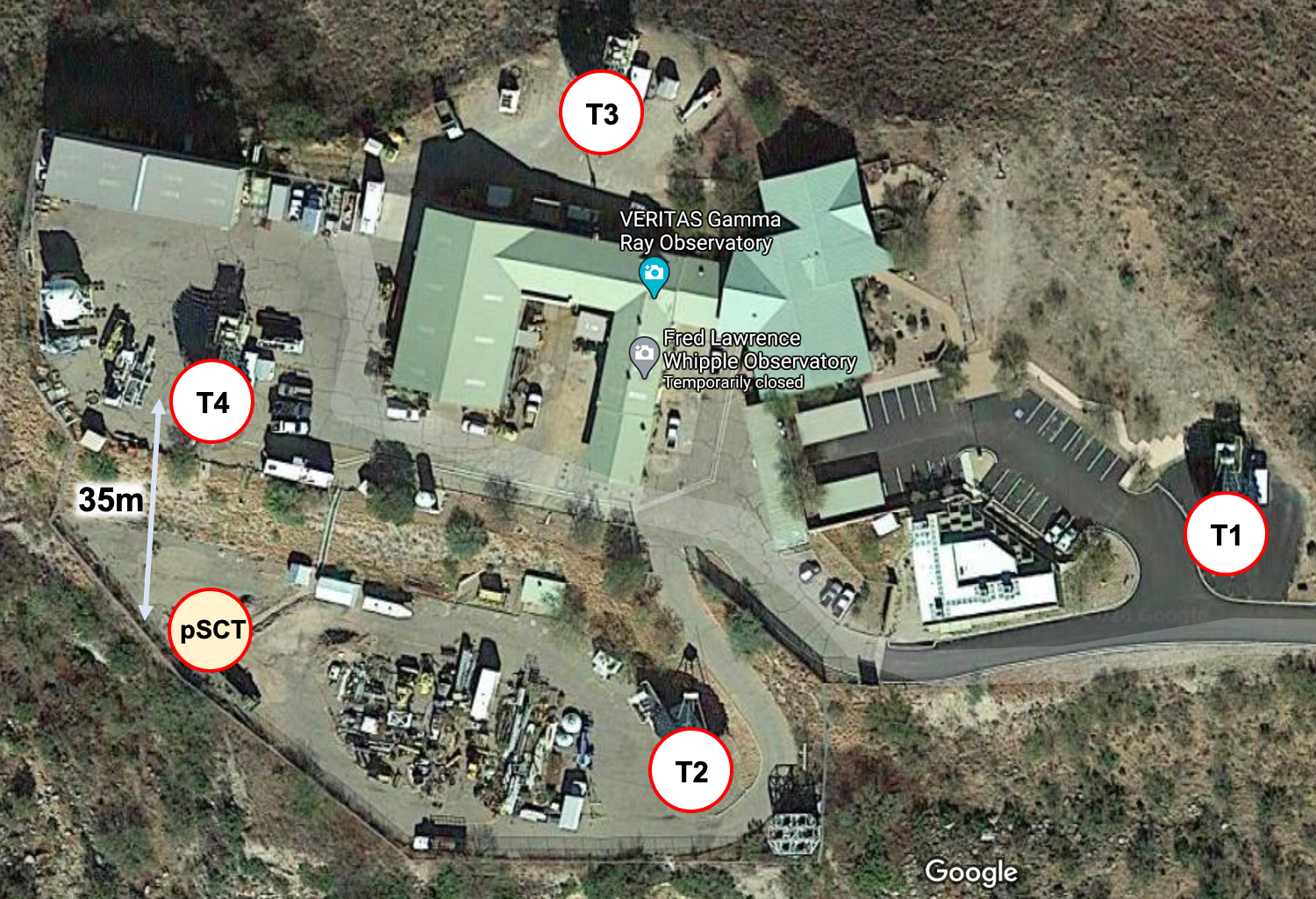}
    \caption{\textbf{Left:} The pSCT utilizes a 9.66 m primary mirror, a 5.4 m secondary mirror and a camera placed between them, facing the secondary mirror. \textbf{Right:} The pSCT is located at the Fred Lawrence Whipple Observatory in southern Arizona alongside the VERITAS Array. An ariel view of the pSCT and VERITAS telescope locations is shown. The pSCT and VERITAS telescope-4 are located only 35 m apart.}
    \label{fig:pSCTlocation}
\end{figure}

\begin{figure*}[b]
    \centering
    \includegraphics[width=0.75\linewidth]{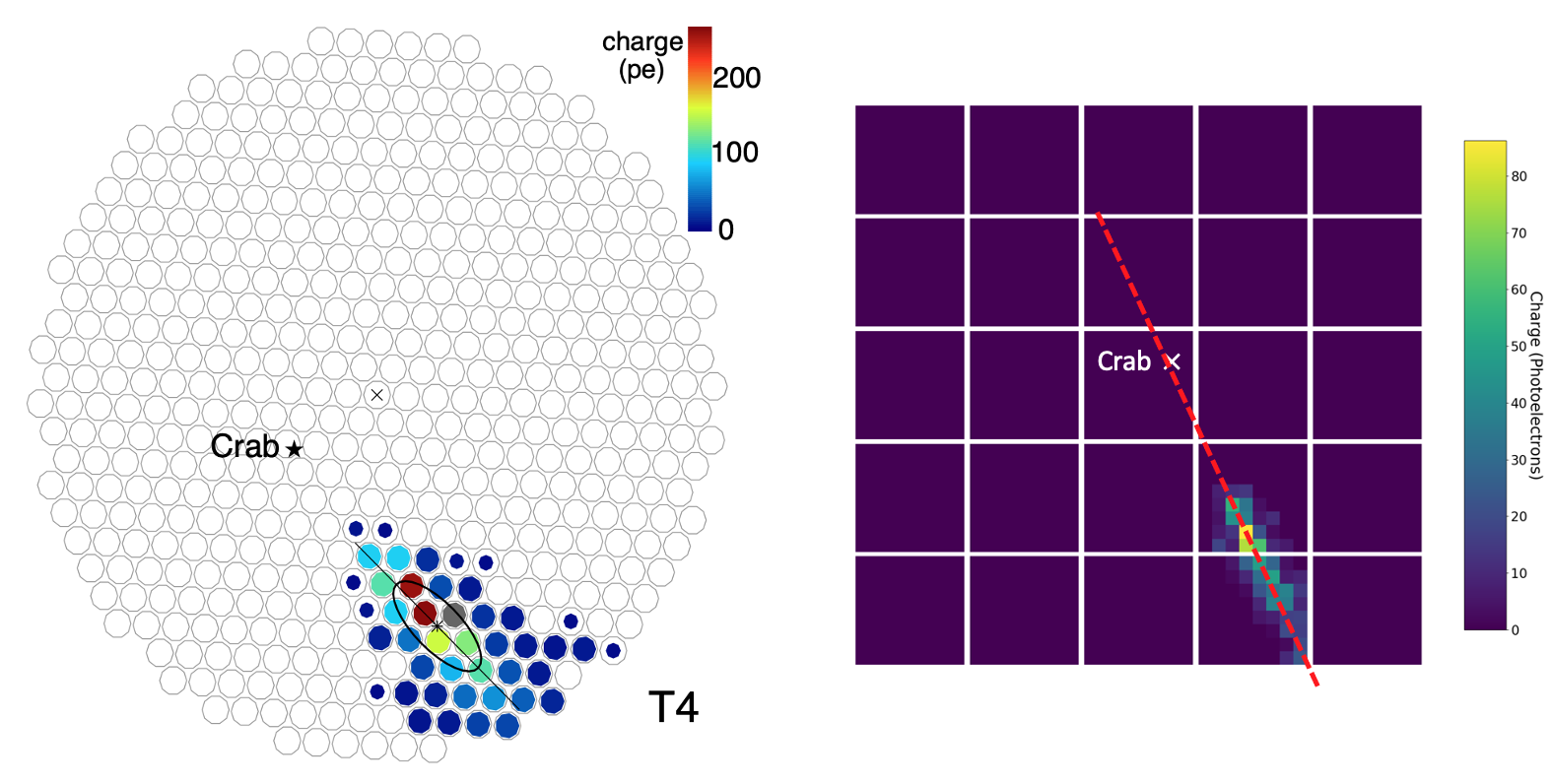}
    \caption{The same air shower event observed by VERITAS telescope-4 (left) and the pSCT(right).}
    \label{fig:AirShower}
\end{figure*}

\begin{figure}
    \centering
    \includegraphics[width=0.6\linewidth]{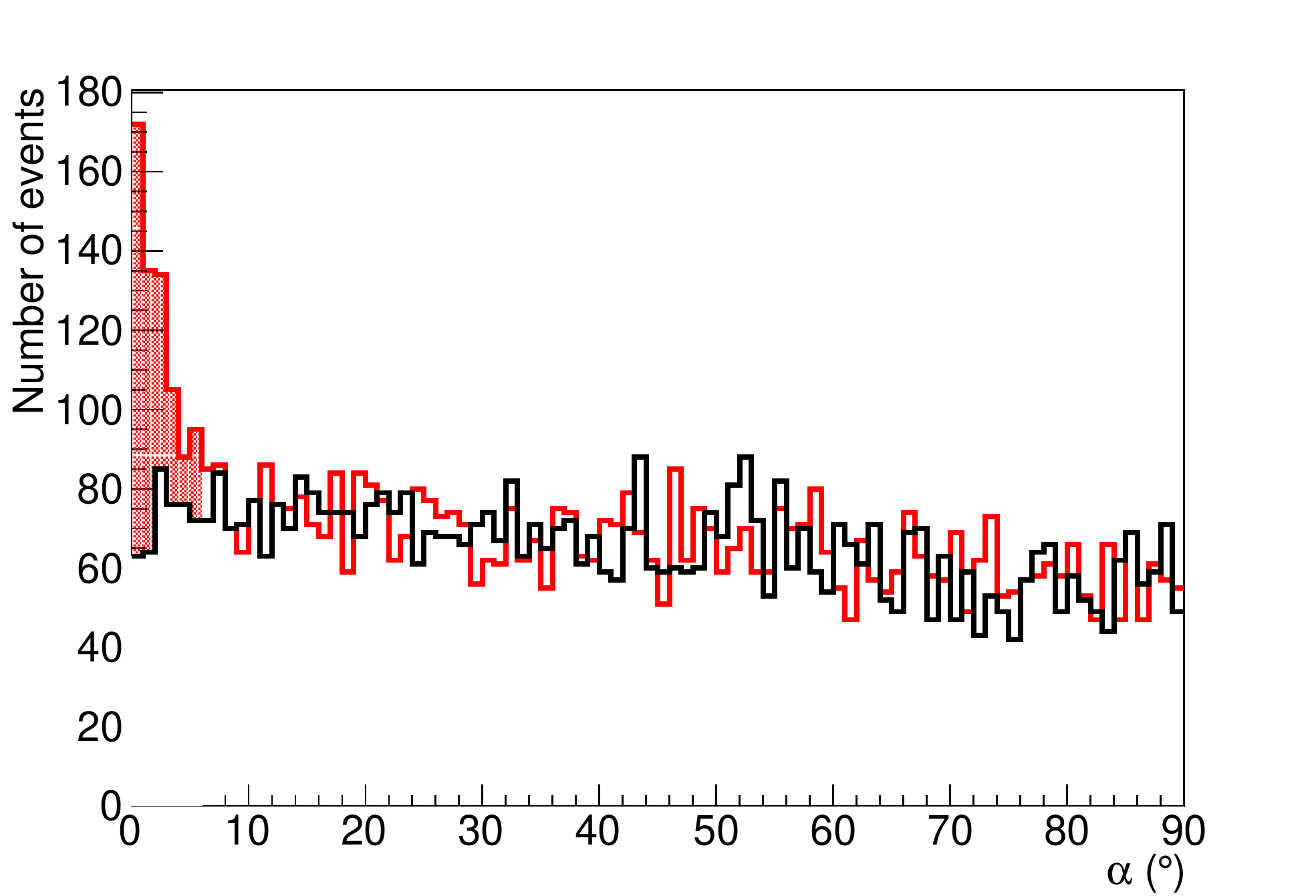}
    \caption{$\alpha$: the angle between the major axis of the event image, and a line joining the centroid of the image to the location of the Crab Nebula. The red histogram is for 17.6 hours of ON source observations, black is for the same duration of OFF source observations, after applying the gamma-ray selection cuts. The shaded region indicates the $<6^{\circ}$ cut on $\alpha$ itself. \cite{adams2021detection}}
    \label{fig:crab_alpha}
\end{figure}

The pSCT is currently undergoing an upgrade consisting of updates to the camera back-end electronics, the camera front-end electronics, and the module SiPMs. In addition, the pSCT focal plane will be fully populated, increasing the number of moduole to 177, the number of pixels to 11,328, and the field of view to 8$^{\circ}$ diameter.

The backplane PCB will be redesigned as the current version is designed to accomodate 32 modules rather than the required 25. The increased number of modules will require 9 backplanes. Instead of a mixture of photosensors all upgraded modules will use third-generation FBK NUV-HD SiPMs. These sensors are a substantial improvement over those currently installed, including lower temperature dependence, improved photon detection, and a lower breakdown voltage. The FEEs will also be upgraded, incorporating next-generation TARGETC ASICs for digitization and sampling as well as a separate TARGET 5 Trigger Extension ASIC for triggering. This new design is expected to significantly reduce trigger noise. \cite{tibaldo2015target} Finally, a custom ASIC called the SMART chip has been developed in order to connect the FEE and SiPMs. The SMART chip will provide improved control over SiPM bias voltage and improved amplification and pulse shaping at the SiPM sensors. Figure \ref{fig:ChargeSpectrum} compares charge spectrums from current and upgraded modules, showing lower noise and significant improvement in charge spectrum resolution. Also included in the upgrade are various auxiliary systems: the heat management system, shutter, and flashers.

\section{Crab Detection}

The pSCT was inaugurated in January of 2019 and the first campaign of observations was conducted in January and February of 2020. Data were primarily taken using an alternating ON/OFF method where 29 minute observations of the source (ON) were taken directly before or after 29 minute observations of an offset field (OFF). Offset fields were chosen such that they covered the same elevation angles as their corresponding ON source observations. A total of 21.6 hours of ON source observations and 17.6 OFF source observations were taken. When possible ON source observations were taken simultaneously with VERITAS. 

The typical trigger rate during these observations was approximately 100 Hz with a majority of these due to electronic noise and approximately 10 Hz artificially injected by uniform illumination of the camera using LED flashers. Events which do not contain an air shower were identified and removed by requiring that at least four adjacent pixels had signals greater than 2 photo-electrons. Remaining events were then cleaned and parameterized using a simple geometrical moment analysis and the Hillas image parameters (size, length, width, distance, $\alpha$, etc.) were determined. \cite{adams2021detection}

The pSCT and VERITAS are co-located at the Fred Lawrence Whipple Observatory in southern Arizona (see figure \ref{fig:pSCTlocation}). This close proximity meant that VERITAS could provide independent information about air showers which both instruments observed simultaneously. 2.2 hours of simultaneous observations resulted in 18 gamma-ray events and 11597 cosmic-ray events recorded by both instruments. Figure \ref{fig:AirShower} shows one such gamma-ray event. The events were classified by VERITAS and identified in pSCT data via timing coincidence. Using only the 2.2 hours of coincident observation, gamma-ray selection criteria were established with the aim of retaining 95\% of the gamma-ray sample.\cite{adams2021detection} 

A separate set of pSCT-only ON/OFF observations were then used for the analysis. The selection cuts were applied to this pSCT-only data. The distribution of the $\alpha$ parameter is shown in Figure \ref{fig:crab_alpha}. There is an excess in ON-source observations at low values of $\alpha$ which corresponds to a statistical significance of $8.6 \sigma$.\cite{adams2021detection} 

\bibliographystyle{JHEP}
\bibliography{references}

\end{document}